\documentclass[%
twocolumn,
superscriptaddress,
 amsmath,amssymb,
 aps
]{revtex4-1}

\usepackage{graphicx}
\usepackage{dcolumn}
\usepackage{bm}
\usepackage{graphicx}
\usepackage{amssymb,amsmath}
\usepackage{float}
\usepackage{longtable}
\usepackage{threeparttable}
\usepackage[toc,page]{appendix}
\usepackage{xcolor}
\makeindex

\begin{document}

\title{Ionic activity in concentrated electrolytes: solvent structure effect revisited}

\author{Amir Levy}
\email{Email: amirlevy@mit.edu}
\affiliation{Departments of Physics, Massachusetts Institute of Technology, Cambridge, Massachusetts 02139 USA}

\author{Martin Z. Bazant}
\email{Email: bazant@mit.edu}
\affiliation{Departments of Chemical Engineering and Mathematics, Massachusetts Institute of Technology, Cambridge, Massachusetts 02139 USA}

\author{Alexei A. Kornyshev}
\email{Email: a.kornyshev@imperial.ac.uk}
\affiliation{Department of Chemistry, Imperial College London, Molecular Sciences Research Hub, White City Campus, London W12 0BZ, and
Thomas Young Centre for Theory and Simulation of Materials, Imperial College London, South Kensington Campus, London SW7 2AZ, United Kingdom}


\begin{abstract}
We revisit the role of the local solvent structure on the activity coefficient of electrolytes with a general non-local dielectric function approach. We treat the concentrated electrolyte as a dielectric medium and suggest an interpolated formula for the dielectric response. The pure water limit is calibrated based on MD simulations and experimental data. Solving our model around a central ion, we find strong over-screening and oscillations in the potential, which are absent in the standard "primitive model" predictions. We obtain mathematically tractable closed-form expressions for the activity coefficients and show reasonable agreement with experimental data. 
\end{abstract}

\maketitle

\newcommand{\abf}{\mathbf{a}}
\newcommand{\pbf}{\mathbf{p}}
\newcommand{\kbf}{\mathbf{k}}
\newcommand{\qbf}{\mathbf{q}}
\newcommand{\xbf}{\mathbf{x}}
\newcommand{\rbf}{\mathbf{r}}
\newcommand{\ubf}{\mathbf{u}}
\newcommand{\dxx}{{\rm d}^3{\bf x}}
\newcommand{\drr}{{\rm d}^3{\bf r}}
\newcommand{\dpp}{{\rm d}^3{\bf p}}
\newcommand{\dqq}{{\rm d}^3{\bf q}}
\newcommand{\kbt}{k_{\rm B} T}

\section{Introduction}

The activity coefficients of concentrated aqueous solutions play an important role in different biological and electrochemical systems\cite{NewmanJohnandThomas-Alyea2012}, and indeed, many models to describe the ionic activity have been proposed over the years. In the original Debye and Hückel (DH) paper from 1923\cite{Debye1923}, the activity coefficient was calculated based on a linearized version of the Poisson-Boltzmann (PB) equation. This resulted in the well-known DH equation, which for symmetric binary electrolytes reads:
\begin{equation}
    \label{OriginalDH}
    \ln \gamma = -\frac{A |z_+z_-|\sqrt{I}}{1+Ba\sqrt{I}},
\end{equation}
where $\gamma$ is the activity coefficient, $z_\pm$ are the valencies of the ions, $I=\sum_{i=\pm} c_i z_i^2 $ is the ionic strength, $c_\pm$ are the ionic concentrations and $a$ is an effective distance of closest approach, roughly equal to the ionic diameter. $A$ and $B$ are constant values that depend on the temperature ($\kbt$), the dielectric constant of the medium ($\varepsilon$) and the unit charge ($e$):
\begin{equation}
    B=\sqrt{\frac{8\pi e^2}{\varepsilon \kbt }}, \,\,\, A= \frac{e^2 B}{2\varepsilon\kbt }.
\end{equation}
The DH equation works well for very dilute electrolytes but fails to even qualitatively capture the activity behavior at higher concentrations. In a following work\cite{Huckel1925}, Huckel added an important term for the activity: the change in self-energy due to variations in the dielectric constant. Experiments measuring the static dielectric constant of ionic solutions were not available at the time, so the proposed model treated the dielectric constant as a fitting parameter. Assuming the dielectric constant of bulk water, $\varepsilon_{\rm Bulk}$, is decreased proportionally to the ionic concentration $c$ ($\varepsilon \approx \varepsilon_{\rm Bulk} - \delta c$), the correction to the DH equation is a simple linear term in concentration. Remarkably, fitting this model to existing activity data actually estimated the dielectric decrement close to measured values, an observation first noted by Hasted et al\cite{Hasted1948} in their paper on the dielectric properties of ionic solutions. 

A linear correction for the DH equation also emerges when considering short-range repulsive forces, via a virial expansion. The virial expansion offers a systematic way to include even higher order terms in concentration. First suggested by Guggenheim\cite{Guggenheim1935}, and further developed by Pitzer\cite{Pitzer1973,Pitzer1974}, accounting for the second and third virial coefficients leads to a very powerful description of the activity. The Pitzer formula, which is essentially the regular Debye-Huckel with corrections to second order in the concentration, is in excellent experimental agreement for hundreds of compounds\cite{Kim1988}. To achieve its high accuracy, the Pitzer model hence requires several fitting parameters: the virial coefficients are not derived from first principles, and the ionic radii are empirical parameters as well. Closely related models were subsequently derived by Bromley\cite{Bromley1973}, Meissner\cite{meissner1972activity} and Chen\cite{Chen1979}.

In the past half a century many more models have been developed on the basis of integral equation approach to statistical theory of fluids, adapted for charged fluids. The Hyper-Netted Chain approximation (HNC) and the Mean Spherical Approximations (MSA) are examples for microscopic derivations of the activity coefficient\cite{Simonin1998,Caccamo1996,SlothPeter1990}. Assuming a hard-sphere repulsion in addition to the Coulombic attraction, the integral equation theories give an approximated way to calculate the pair correlation function between any two ions. Usually, numerical methods are required to solve the integral equations. The activity is expressed in terms of the correlation function, without a simple closed-form formula. Another drawback of the integral equation model is that they too require some fitting parameters.   

Not going into a comparison of these different approaches, we only stress that they where all derived for the primitive model of the solvent. Ions interact there via Coulomb law like they would if the solvent was a dielectric continuum, with a macroscopic dielectric constant. At the same time we know from molecular simulations that in polar solvents, water, in particular, the potential of mean-force between the ions exhibit decaying oscillations with the periodicity of the order of the diameter of the solvent molecules, with signatures of overscreening effect, and only at long distances it would approach the macroscopic Coulomb interactions, as a limiting law. How this fact would reveal itself in thermodynamics of electrolytes? One way to answer this question would be to incorporate the effects of the molecular structure of the solvent via replacing the Coulomb pair interaction potential in the above-mentioned approaches with the correspondingly modified ones. Alternatively, one could incorporate the differences from the primitive Coulomb into the short-range part of the interaction potential. Such short-range part would then extend few times farther than the average diameter of ions. In order to justify such efforts, we will do here something yet simpler:  We will combine the Debye-Hueckel approach with a nonlocal electrostatic description of the solvent. Although such approach will not take into account complex correlations in a concentrated electrolyte, it will be a step towards connecting the correlations of the bound charge density of the solvent subsystem (molecular correlations) and the ion-ion correlations in the electrolyte plasma.   Such an approach will work as an interpolation. Following this root, we will result in a closed-form expression which as we will see will describe the behavior of activity coefficients very well. 

Such an approach has been, actually, proposed and tried long ago \cite{holub1976polar,kornyshev1996shape}. We revisit it below, showing that for the updated approximation of the form of the nonlocal dielectric function of a pure solvent that qualitatively reproduces the simulation results for water \cite{bopp1996static,bopp1998frequency} we can obtain very reasonable results for the activity coefficients and explore certain trends in their dependence on electrolyte concentration.

\section{Model}

Our goal is to build a phenomenological description of the dielectric function of ionic solutions, that accounts for both the solvent molecules and the ions and would enable us to calculate the ionic activity coefficient. In a constant dielectric medium (the so-called "primitive" model), one can derive the dielectric response directly from the Poisson-Boltzmann equation. However, a constant dielectric medium is an approximation suitable for large ion-ion separations. At shorter separations, the molecular ordering of the water gives rise to a complicated dielectric response. Empirical formulae for the dielectric function have been suggested in the literature\cite{kornyshev1996shape, kornyshev1997overscreening}, in relation with computer simulation results \cite{bopp1996static} and experimental data\cite{soper1994orientational}. We will now show how we extend the pure-water empirical dielectric response for ionic solutions, by building an interpolated function that satisfies the limiting behaviors. 

Within linear nonlocal electrostatics, electrical induction and electric field are related by nonlocal
constitutive relation: $D_{\alpha}(\mathbf{r})=\sum_{\beta} \int d \mathbf{r} \varepsilon_{\alpha \beta}\left(\mathbf{r}-\mathbf{r}^{\prime}\right) E_{\beta}\left(\mathbf{r}^{\prime}\right)$, where $\varepsilon_{\alpha \beta}\left(\mathbf{r}-\mathbf{r}^{\prime}\right)$ is the nonlocal dielectric tensor. In the macroscopic electrostatics $\varepsilon_{\alpha \beta}\left(\mathbf{r}-\mathbf{r}^{\prime}\right)=\varepsilon \delta_{\alpha \beta} \delta\left(\mathbf{r}-\mathbf{r}^{\prime}\right)$ which reduces the constitutive relation to the common $\mathbf{D}(\mathbf{r})=\varepsilon \mathbf{E}(\mathbf{r})$. All information about the correlations of the
bound charge density in the medium are contained in the form of the tensor $\varepsilon_{\alpha \beta}\left(\mathbf{r}-\mathbf{r}^{\prime}\right)$. Referring the reader to Ref.\cite{kornyshev1981nonlocal} for details, we mention that in homogeneous and isotropic media, electrostatic equations will be conveniently expressed through the Fourier transform of this tensor $\tilde{\varepsilon}_{\alpha \beta}(\mathbf{k})$, and more precisely through its longitudinal component $\widetilde{\varepsilon}_{ \|}(k)=\sum_{\alpha \beta} \frac{k_{\alpha} k_{\beta}}{k^{2}} \tilde{\varepsilon}_{\alpha \beta}(\mathbf{k})$, often called simply $\tilde{\varepsilon}(k)$ . Long wave-length limit (small $k$) recovers macroscopic behaviour, large $k$ , probes short range correlations. For instance, speaking about pure solvent $k \sim 2 \pi / d$ , where $d$ is diameter of water molecule, would characterize the molecular packing effects. For much larger, $\tilde{\varepsilon}(k)$ will
approach short range dielectric constant due to electronic polarizability of the molecules. We denote the corresponding dielectric constant in that limit $\varepsilon_*$.

In the long wavelength limit, the Poisson-Boltzmann equation for a binary monovalent solution reads:
\begin{eqnarray}
    \varepsilon_{\rm Bulk} \nabla^2 \phi(\rbf) = 8 \pi e c \sinh [e \beta \phi(\rbf)]-4\pi \rho_{\rm ext}(\rbf),
\end{eqnarray}
where $c$ is the bulk ionic concentration, $\beta = 1/k_{\rm B} T$ is the inverse temperature and $\rho_{\rm ext}$ is an external charge distribution. In the linear (DH) regime, the PB equation is a second order differential equation, or an algebraic equation in Fourier space:
\begin{eqnarray}
\label{pb_k}
    \varepsilon_{\rm Bulk} k^2\left[1 + \frac{8\pi c  e^2 \beta}{\varepsilon_{\rm Bulk} k^2}\right] \tilde\phi(\kbf)=4\pi\tilde\rho_{\rm ext}(\kbf).
\end{eqnarray}
Comparing Eq.~\ref{pb_k} to the Poisson equation, we can immediately write the dielectric response of ionic solutions in the limit of large wavelengths:
\begin{eqnarray}
\label{eps_ion_bulk}
\tilde\varepsilon_{\rm c,bulk}(\kbf) = \varepsilon_{\rm Bulk}\left[1+\frac{1}{(k\lambda_D)^2}\right],
\end{eqnarray}
where $\lambda_{\rm D} = (8\pi c  e^2 \beta/\varepsilon_{\rm Bulk})^{-1/2}$ is the Debye screening length. The divergence at small wave-numbers corresponds to the screening of the potential at distances larger than the Debye screening length. At smaller distances the screening effect is negligible, and dielectric response is only influenced by the water. This will remain true even if we consider a more complicated expression for the water dielectric response, rather than $\varepsilon_{\rm Bulk}$. Hence, we can write a simple interpolated formula for the dielectric response by replacing  $\varepsilon_{\rm Bulk}$ with the full ${\tilde \varepsilon_w}(k)$:
\begin{eqnarray}
\label{epsC}
{\tilde \varepsilon}_c(\kbf) = {\tilde \varepsilon_w}(k)\left[1+\frac{1}{k^2\lambda_D^2}\right]
\end{eqnarray}
This approach is similar to interpolation implemented in Refs.\cite{kornyshev1981nonlocal,kornyshev1983non} in terms of the limiting cases covered, but its form is slightly different, reflecting stronger coupling between the solvent structure and the ionic screening. 
Note that this interpolated response satisfies both the long and short wavelength limits. In the long-wavelength ${\tilde \varepsilon_w}(k)\rightarrow \varepsilon_{\rm Bulk}$ and we recover Eq.~\ref{eps_ion_bulk}. In the short-wavelength, the ionic contribution is neglected as we recover the pure-water response.  

By design, the interpolated formula is expected to work well if there is a separation of length-scales, and the Debye length is much larger than the molecular size of the solvent. In this limit, however, the predicted ionic activity will coincide with classical DH theory. Interesting physics emerges as we increase the concentration, and enter a regime where both ions and water molecules play a major role. 

Within the linear approximation, this interpolation formula for the dielectric constant provides all the necessary information required to derive the activity coefficient. Let us now, following Ref.17 (first time derived in \cite{holub1976polar}), use the charging process to evaluate the activity of ions, by considering a spherical particle immersed in a dielectric medium. By slowly turning on the charge, the energy is determined by the potential at the surface of the ion:
\begin{eqnarray}
\label{uDef}
u=\int_0^e dq \phi_q(r=a), 
\end{eqnarray}
where $\phi_q(r)$ is the electrostatic potential around a charged particle with charge $q$, and $a$ is the effective radius of the sphere, related to the distance of closest approach to the ion. A simple way of estimating the electrostatic potential is by letting water to permeate the ion, and solving the Poisson equation in k-space:
\begin{equation}
\label{phiFourier}
    \phi_q(r)=\int \frac{d\kbf}{(2\pi)^3} \frac{4\pi q{\tilde \rho}(\kbf)}{{\tilde \varepsilon}_c(\kbf)k^2} {\rm e}^{i \kbf \cdot \rbf}=\frac{2}{\pi}\int_0^\infty dk \frac{\sin(kr)}{kr}\frac{q{\tilde \rho}(k)}{\tilde\varepsilon_c(k)},
\end{equation}
where ${\tilde \rho(\kbf)}$ is the Fourier transform of the charge distribution, called also an ionic form-factor:
\begin{equation}
  \tilde{\rho}(k)  =\int d\rbf \rho(\rbf) {\rm e}^{-i \kbf \cdot \rbf}.
\end{equation}
Following Ref.~\cite{kornyshev1996shape}, we use a smeared charge distribution, defined as follows:
\begin{eqnarray}
{\tilde{\rho}(k)=\frac{1}{\eta\left(a^{2}+\eta^{2}\left(2-e^{-a / \eta}\right)\right)}\cdot\left\{\frac{\eta a \sin k a}{k\left(1+\eta^{2} k^{2}\right)}\right.} \nonumber\\ {+\frac{\eta^{3}\left(2 \cos k a-e^{-a / \eta}\right)}{\left(1+\eta^{2} k^{2}\right)^{2}} \}},\nonumber\\
\end{eqnarray}
where $\eta$ is the smearing parameter, which describes the width of the ionic charge shell; for $\eta\rightarrow0$ the form-factor reduces to the Aschcroft form: $\tilde{\rho}(k)=\sin(ka)/ka$. Combining Eqs.~(\ref{uDef}) and (\ref{phiFourier}) we obtain:
\begin{eqnarray}
u&=&  \frac{e^2}{\pi}\int_0^\infty dk \frac{\tilde \rho(k)}{{\tilde \varepsilon}_c(k)} \frac{\sin(ak)}{ak}.
\end{eqnarray}
The excess chemical potential of moving an ion from bulk water to ionic solution with concentration $c$, is given by (in units of thermal energy, $\kbt$):
\begin{eqnarray}
\label{lnGamma1}
\ln \gamma &=& \beta\left[u(c)-u(c=0)\right]
\nonumber\\
& = & \frac{l_{\rm B}}{\pi}\int_0^\infty dk  \frac{\sin^2(ak)}{(ak)^2}\left[\frac{\tilde \rho(k)}{{\tilde \varepsilon}_{\rm c}(k)}-\frac{\tilde \rho(k)}{{\tilde \varepsilon}_{\rm c=0}(k)}\right],\nonumber\\
\end{eqnarray}
where $l_{\rm B} = \beta e^2$ is the vacuum Bjerrum length. Using the interpolated formula for $\varepsilon_c(k)$ (Eq.~\ref{epsC}), we can write the chemical potential in terms of the water dielectric constant and relate it to the DH limiting law: 
\begin{eqnarray}
\label{lnGamma2}
\frac{\ln \gamma}{\ln \gamma^{\rm DH}} =  \frac{2}{\pi} \int_0^\infty \frac{dk}{\lambda_D}   \frac{\sin(ak)}{ak} \frac{\varepsilon_{\rm bulk}}{\tilde{\varepsilon}_w(k)}\frac{\tilde{\rho}(k)}{k^2 + \lambda_D^{-2}},
\end{eqnarray}
where $\ln\gamma^{\rm DH}$ is the classical activity formula (in the limit of $a \rightarrow 0$):
\begin{eqnarray}
\ln\gamma^{\rm DH} = -\frac{l_{\rm B}}{2\varepsilon_{\rm bulk} \lambda_D }.
\end{eqnarray}
Finally, we need to suggest a model for the solvent dielectric function. So far we have only specified the limits it must hold: it equals bulk values ($\varepsilon \approx 80 $) at small wave-vectors and some small value $\varepsilon_*$ at large ones. It is instructive to introduce a weighting function $f(k)$, that equals $1$ at the large wavelength limit, and $0$ for short wavelengths, so we can write a general dielectric function as:
\begin{eqnarray}
{\tilde \varepsilon}_w(k)=\left[(\varepsilon_*)^{-1}+\left(\varepsilon_{\rm bulk}^{-1}-(\varepsilon_*)^{-1}\right)f(k)\right]^{-1}. 
\end{eqnarray}
A simple $f(k)$ that satisfies the corrects limits is a Lorentzian shape:
\begin{eqnarray}
f(k)=\frac{1}{1+k^2\Lambda^2}.
\end{eqnarray}
The Lorentzian shape captures some effects of solvent structure at long wave-length, implying that water molecules are correlated, and their correlation is exponentially decreasing with a decay length $\Lambda$. But it misses to correctly describe the short range behaviour: molecular dynamics simulations of water molecule reveal a much more complicated structures with $k$-dependence reflecting resonance effects of over-screening \cite{bopp1996static,kornyshev1997nonlocal,fedorov2007unravelling}. In the spirit of work \cite{kornyshev1996shape} we could account for over-screening by the following formula for $f(k)$:
\begin{eqnarray}
\label{fOverScreening}
f(k)=\frac{ (1+(\Lambda^2 Q^2))^2}{(1+(k\Lambda - Q\Lambda)^2) (1+(k\Lambda + Q\Lambda)^2)},
\end{eqnarray}
where $\Lambda$ describes the correlation length as before, and $Q\approx2\pi/d_w$ is the wavelength for oscillations, which is determined by the molecular size of water. In this work, however, we propose a more general form to better describe the permittivity in the intermediate wave-numbers range:
\begin{eqnarray}
\label{fOverScreening}
f(k)=\frac{\alpha}{(1+\Lambda^2 k^2)^2} + \frac{(1-\alpha)(1+(\Lambda^2 Q^2))^2}{(1+(k\Lambda - Q\Lambda)^2) (1+(k\Lambda + Q\Lambda)^2)}.
\nonumber\\
\end{eqnarray}
It mimics the basic features of the response function as found in \cite{bopp1996static}, approved by experimental data (\cite{soper1994orientational}).  Our hybrid model is illustrated in Fig.~1 by looking at the response function $\chi(k)=1/\varepsilon_*-1/{\tilde \varepsilon}_w(k)$. The large peak around $k=3{\rm \AA}^{-1}$ is related to overscreening and will lead to oscillations with a period just below the molecular diameter. The hybrid model corrects the longer range behavior of the overscreening model, where the dielectric function is expected to be slightly reduced, similar to the Lorentizian model as obtained in Ref. \cite{bopp1996static}. 

\begin{figure}[h]
  \includegraphics[width=\columnwidth ]{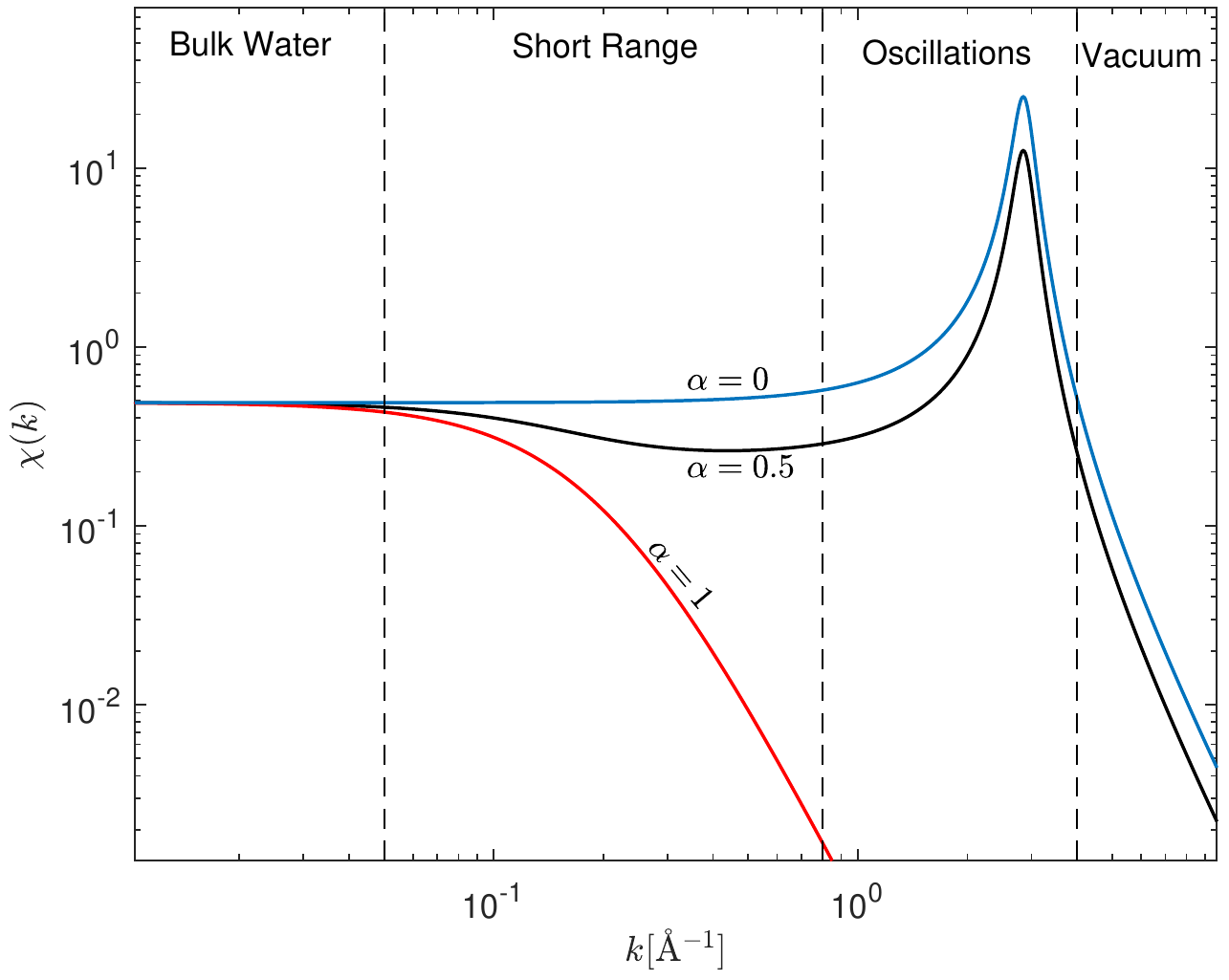}
  \caption{The response function $\chi(k)$ as a function of wavelength, for three models of dielectric functions: simple Loretntizan model, over-screening model and a hybrid model.  }
  \label{fig1}
\end{figure}

\section{Results}
Let us now calculate the activity coefficient for a typical ionic solution, for concentrations ranging from the very dilute to moderately concentrated, using Eq.~\ref{lnGamma1}. The parameters for the water dielectric response that we adopt here are summarized in Table.~1. Most of them were determined according to previous studies of pure water, however the hybrid-model weighting function $\alpha$ was fitted to experimental activity data. 
\begin{table}[]
\begin{tabular}{l|l|l}
\hline
\textbf{parameter}             & \textbf{symbol}          & \textbf{value} \\ \hline \hline
Short wavelength permittivity            & $\varepsilon_*$          & $2$            \\ 
Solvent bulk permittivity      & $\varepsilon_{\rm Bulk}$ & $80$           \\ 
Solvent oscillation wavelength & $Q$             & $\frac{2\pi}{2.2}$\AA$^{-1}$               \\ 
Solvent correlation length           & $\Lambda$                & $5$\AA         \\ 
Weight parameter               & $\alpha$                  & $0.03$          \\ 
Smearing parameter             &$\eta$                      &$0.5$\AA         \\ \hline
\end{tabular}
 \caption{Parameters for our hybrid model of pure-water dielectric response (Eq.~\ref{fOverScreening}). }
\end{table}

To better understand the activity coefficient, we first examine the potential profile around a spherical ion. Fig.~2 shows the potential for increasing ionic concentrations, compared with standard DH approximation (Fig.~2 inset). The results are based on a distance of closest approach (ionic diameter, $a$) of $3.5$\AA, and exemplify how the non-local permittivity completely changes the potential profile and leads to a non-linear concentration dependence. 
As we increase the concentration, the DH screening cloud gets narrower, and the potential is strongly screened. In contrast, the oscillating structure, predicted by the non-local dielectric model, persists even in high molalities. 
\begin{figure}[h]
  \includegraphics[width=\columnwidth ]{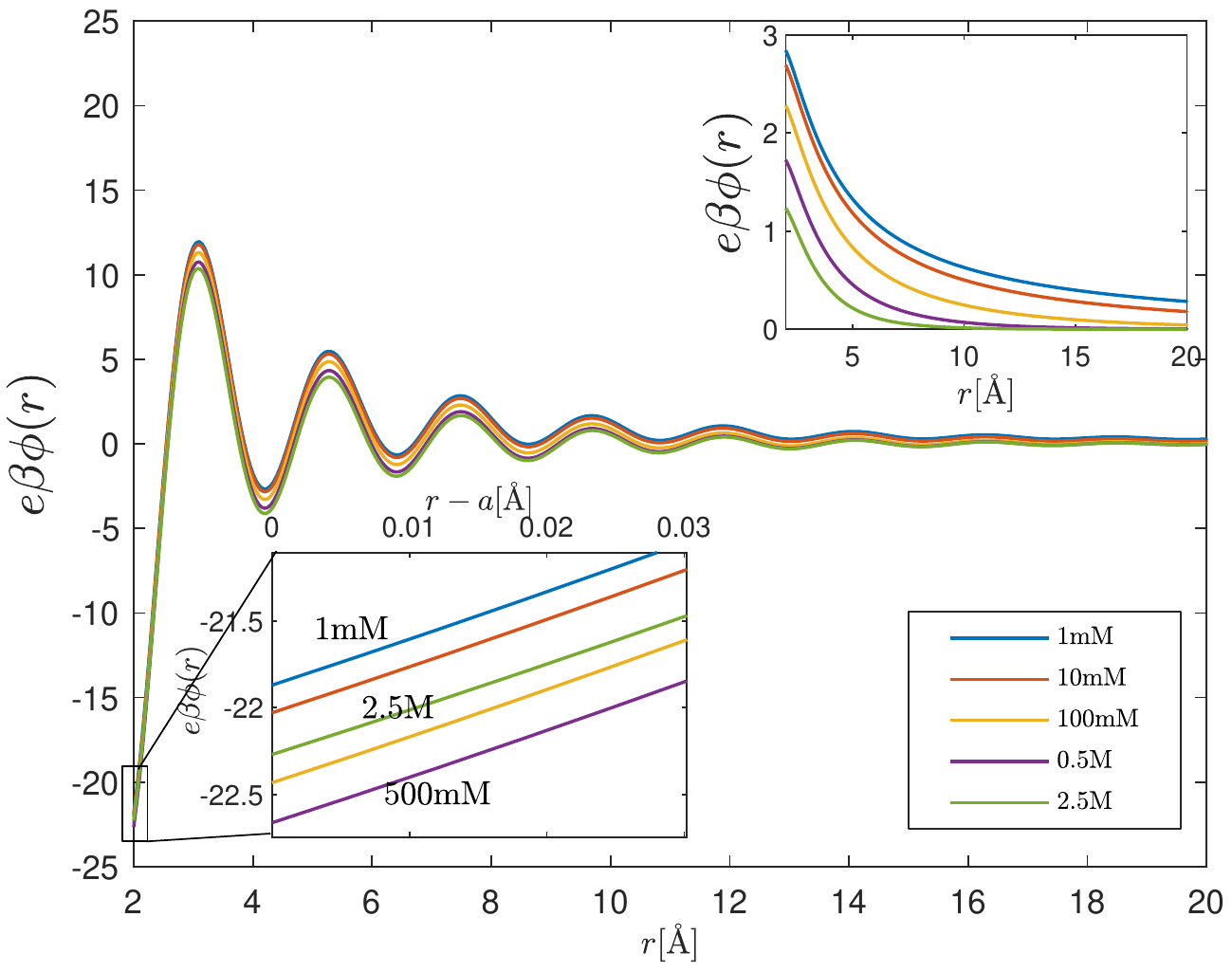}
  \caption{The dimensionless electrostatic potential, $e\beta\phi(r)$, around a spherical ion, for ionic concentrations $1$mM, $10$mM, $100$mM, $0.5M$ and $2.5$M, based on a the non-local dielectric function described by Eq.~\ref{epsC}. Parameters of the non-local function are given in Table.~1. The ionic diameter used is $3.5$\AA. Right inset- the dimensionless electrostatic potential in a constant dielectric medium. Left inset- the dimensionless electrostatic potential profile near the surface of the sphere. At larger distances fro the ion, as well as at any distances for the case of constant permittivity, the potential decreases with electrolyte concentration, but in the vicinity of the ion the effect is non-monotonic.   }
  \label{fig1}
\end{figure}

From the charging process, we know that the activity is related to the potential at the surface of the charged ion. Two competing effects determine this potential for non-local dielectrics. For small ionic concentrations, the potential is lowered, as a result of the interaction with the screening cloud. This change allows us to recover the DH limiting law, as expected. We note, however, that in contrast to a constant-dielectric picture, the potential itself is negative, and increases in magnitude. As the ionic concentration increases, the amplitude of the oscillations is reduced, which leads to an opposite trend: the screening of the oscillations results in a smaller magnitude potential, i.e., it becomes less negative. 

The resulting activity profiles are shown in Fig.~3, for three ionic diameters ($a=2$\AA, $3{\rm \AA}$ and $4$\AA). For comparison, three experimentally measured curves of activity coefficients are shown as well. The experimental data were taken from ref \cite{Bromley1973} and corresponds to three monovalent ionic solutions: KCl, NaCl, and LiCl, representing three different cation sizes. Qualitatively we see that our model is able to capture the correct trend, including the increased activity at high concentrations, as well as some size dependence of the activity coefficient. The bare cation diameters for the potassium, chloride, and lithium are $1.4$\AA, $1.94{\rm \AA}$ and $2.82{\rm \AA}$, respectively\cite{Marcus1983}, which are only slightly lower than the values we consider here. While we are not claiming this is a complete model, we show that with reasonable parameters, the water structure alone can explain much of the overall shape of the activity vs concentration for different ions, without resorting to correlations or concentration-dependent permittivity.

\begin{figure}[h]
  \includegraphics[width=\columnwidth ]{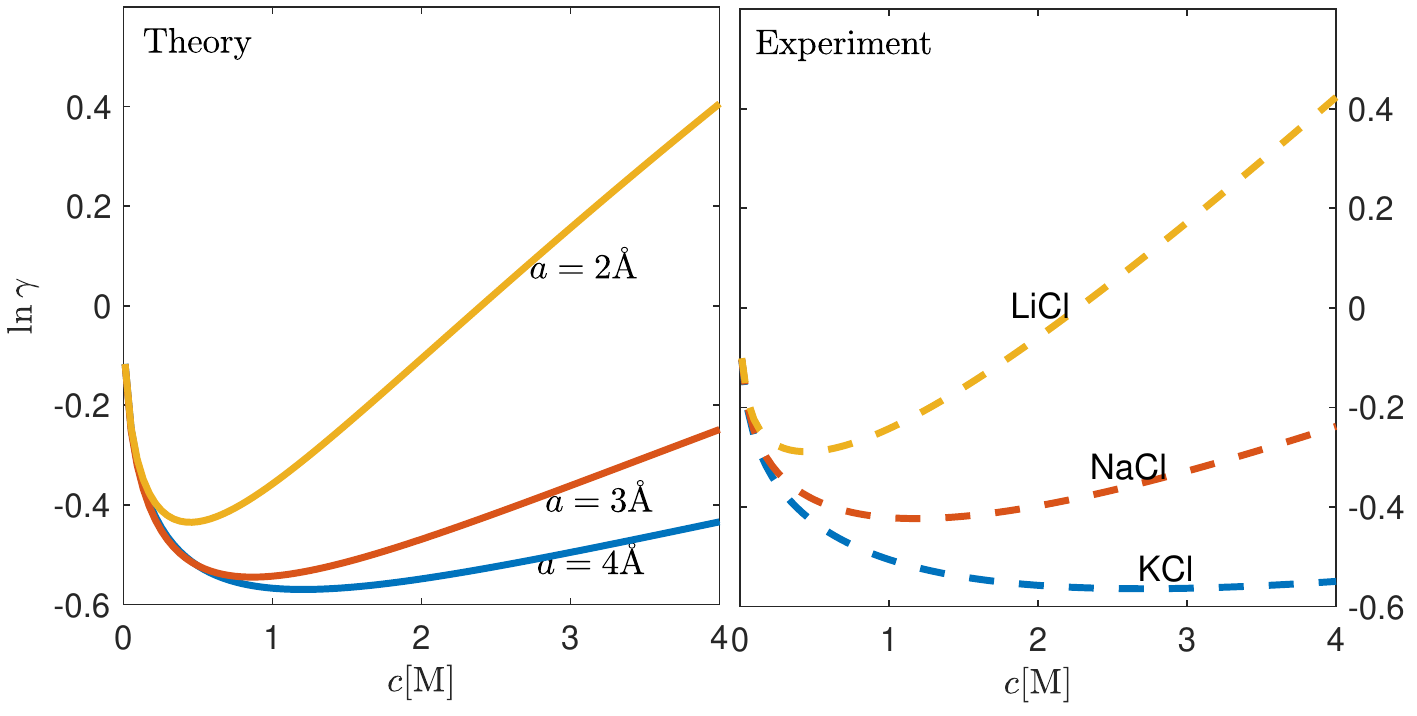}
  \caption{Activity coefficients as a function of ionic concentrations. Left- The activity coefficient for monovalent binary solutions, based on the non-local permittivity model, Eq.~\ref{lnGamma1}. Three different activity curves are shown, corresponding to three ionic diameters (from bottom to top): $4$\AA, $3$\AA and $2$\AA. The parameters of the pure water dielectric function are summarized in Table.~1, and the smearing parameter was taken to be $\eta=0.5$\AA. Left-  Experimental data of activity coefficients for three ionic solutions (from bottom to top): KCl, NaCl and LiCl. Data is taken from \cite{Bromley1973}.}
  \label{fig1}
\end{figure}
\section{Discussion}

The match between the experimental activity coefficient and our model illustrates the importance of the local water structure on ionic activity. Our theory supports the original argument of Huckel himself, as well as several recent papers\cite{Liu2015a,vincze2010nonmonotonic}, that differences in the solvation energy play a central role in determining the activity coefficient. In fact, it is the main source of increasing activity at moderate salt concentrations, reversing the decreasing trend of DH theory for screening at low concentration, even before ion-ion correlations become important at high concentrations. Yet, the interplay between solvent molecules and ions is usually either ignored altogether or artificially added as an additional contribution, based on a  concentration-dependent bulk dielectric constant. By using an independently validated dielectric response of the solvent, we show the significance of the short-range water behavior, that is only vaguely captured by an effective reduced dielectric constant. 

It is important to note that we have neglected several other important effects that are known to play a role in determining the activity coefficient. First, our dielectric function is based on a linearization of the Poisson-Boltzmann equation and, thus,  non-linear effects in the polarization of the ionic atmosphere are neglected. Moreover, extensions to the PB equation, such as ones that account for finite size ions\cite{Bazant2009,Borukhov1997,Bikerman1942,Kilic2007,Kornyshev2007}, are not considered. Size and packing constraints will rapidly increase the activity coefficient when the packing fraction becomes significant. Theories of primitive models in a constant dielectric medium, supported by Monte-Carlo simulations, have shown that both size effects and non-linear contribution can be significant \cite{card1970monte}. Another necessary contribution to the activity comes from ion-ion correlations and is especially pronounced at high concentrations. Such contributions can naturally fit into a non-local dielectric response framework by introducing a correlation length, $l_c$, describing the lowest order correction to the bulk dielectric constant as results of ion-ion interactions: $\varepsilon(k)\approx\varepsilon_{\rm Bulk}(1+l_c^2 k^2)$\cite{Bazant2011}. Interestingly, ion-ion correlations have an opposite sign compared with water-related correlations, as the second-order expansion of the pure water permittivity gives a negative contribution, reducing the permittivity. The corresponding solvent correlation length interpolates between $Q^{-1}$ and $\Lambda$, and simplifies (for $Q\Lambda\gg 1$) to:
\begin{equation}
    |l_c^{\rm solvent}| \approx 2\Lambda \frac{\varepsilon_{\rm Bulk}}{\varepsilon_*}\sqrt{\beta +  \frac{1-\beta}{(Q\Lambda)^2}}.
\end{equation}

Last but not least, the effects of the electric field of ions on water structure have been neglected, as well as disturbance of the structure by their mere presence, which were both shown to be potentially important \cite{kornyshev1997nonlocal,fedorov2007unravelling}. Indeed, the detailed studies of Ref. \cite{fedorov2007unravelling}, based on integral equation approach to the description of molecular correlations in water and molecular dynamic simulations, reveal a complicated dielectric response, with a strong non-linear component at high electric fields and sensitivity to the polarity of the ions. These limitations, as well as other non-electrostatic interactions that were omitted, limit the adequacy of our model. It is therefore expected that with virtually no ion-specific fitting parameters, apart from the effective 'diameter' of the closest approach, our model would only predict the correct trends, and not exact values. Our formula for the dielectric response, Eq.~\ref{epsC}, is only a first step in the right direction. It is the simplest form that recovers the correct behaviors in both the very short and very long wavelength limits. To improve the results, and get a quantitative agreement with experiments, more elaborate models are required.

\bibliography{sample}
\end{document}